\documentclass{aa}
\input epsf

\begin{document}

\title{Search for TeV gamma ray emission from the Andromeda galaxy}

\titlerunning{Search for TeV gamma ray emission from M31}
\authorrunning{F. Aharonian et al.}

\author{
F.A.~Aharonian\inst{1},
A.G.~Akhperjanian\inst{7},
M. Beilicke\inst{4},
K.~Bernl\"ohr\inst{1},
H.~Bojahr\inst{6},
O.~Bolz\inst{1},
H.~B\"orst\inst{5},
T.~Coarasa\inst{2},
J.L.~Contreras\inst{3},
J.~Cortina\inst{2},
S.~Denninghoff\inst{2},
V.~Fonseca\inst{3},
M.~Girma\inst{1},
N.~G\"otting\inst{4},
G.~Heinzelmann\inst{4},
G.~Hermann\inst{1},
A.~Heusler\inst{1},
W.~Hofmann\inst{1},
D.~Horns\inst{1},
I.~Jung\inst{1},
R.~Kankanyan\inst{1,7},
M.~Kestel\inst{2},
J.~Kettler\inst{1},
A.~Kohnle\inst{1},
A.~Konopelko\inst{1},
H.~Kornmeyer\inst{2},
D.~Kranich\inst{2},
H.~Krawczynski\inst{1,}$^\%$,
H.~Lampeitl\inst{1,4},
M. Lopez\inst{3},
E.~Lorenz\inst{2},
F.~Lucarelli\inst{3},
O.~Mang\inst{5},
H.~Meyer\inst{6},
R.~Mirzoyan\inst{2},
A.~Moralejo\inst{3},
E.~Ona\inst{3},
M.~Panter\inst{1},
A.~Plyasheshnikov\inst{1,}$^\S$,
G.~P\"uhlhofer\inst{1},
G.~Rauterberg\inst{5},
R.~Reyes\inst{2},
W.~Rhode\inst{6},
J.~Ripken\inst{4}
A.~R\"ohring\inst{4},
G.P.~Rowell\inst{1},
V.~Sahakian\inst{7},
M.~Samorski\inst{5},
M.~Schilling\inst{5},
M.~Siems\inst{5},
D.~Sobzynska\inst{2,}$^*$,
W.~Stamm\inst{5},
M.~Tluczykont\inst{4},
H.J.~V\"olk\inst{1},
C.A.~Wiedner\inst{1},
W.~Wittek\inst{2}}

\institute{Max Planck Institut f\"ur Kernphysik,
Postfach 103980, D-69029 Heidelberg, Germany \and
Max Planck Institut f\"ur Physik, F\"ohringer Ring
6, D-80805 M\"unchen, Germany \and
Universidad Complutense, Facultad de Ciencias
F\'{i}sicas, Ciudad Universitaria, E-28040 Madrid, Spain
\and
Universit\"at Hamburg, Institut f\"ur
Experimentalphysik, Luruper Chaussee 149,
D-22761 Hamburg, Germany \and
Universit\"at Kiel, Institut f\"ur Experimentelle und Angewandte Physik,
Leibnizstra{\ss}e 15-19, D-24118 Kiel, Germany\and
Universit\"at Wuppertal, Fachbereich Physik,
Gau{\ss}str.20, D-42097 Wuppertal, Germany \and
Yerevan Physics Institute, Alikhanian Br. 2, 375036 Yerevan,
Armenia\\
\hspace*{-4.04mm} $^\S\,$ On leave from
Altai State University, Dimitrov Street 66, 656099 Barnaul, Russia\\
\hspace*{-4.04mm} $^\%$ Now at Yale University, P.O. Box 208101, New Haven, CT 06520-8101, USA\\
\hspace*{-4.04mm} $^*$ Home institute: University Lodz, Poland\\
}

\maketitle


\begin{abstract}
Using the HEGRA system of imaging atmospheric Cherenkov telescopes,
the Andromeda galaxy (M31) was surveyed for TeV gamma ray emission.
Given the large field of view of the HEGRA telescopes, three 
pointings were sufficient to cover all of M31, including also M32 and
NGC205. No indications for point sources of TeV gamma rays were
found. Upper limits are given at a level of a few percent of the
Crab flux. A specific search for monoenergetic gamma-ray lines from
annihilation of supersymmetric dark matter particles accumulating
near the center of M31 resulted in flux limits in the $10^{-13}$~cm$^{-2}$s$^{-1}$
range, well above the predicted MSSM flux levels except for models with
pronounced dark-matter spikes or strongly enhanced annihilation rates.

\end{abstract}

\section{Introduction}

During the last decade, Imaging Atmospheric Cherekov Telescopes
(IACTs) have extended the spectrum of astronomical observations
into the TeV energy regime, and a small number of galactic sources such as
plerions and Supernova remnants as well as extragalactic 
sources -- active galactic nuclei (AGN) -- have been detected and
studied (e.g. Weekes, 2001). In particular with stereoscopic
systems of Cherenkov telescopes, such as the HEGRA telescope
system  (Daum et al. 1997, Aharonian et al. 1999, Konopelko et al. 1999), 
surveys of moderately extended regions of the sky have become
possible (Aharonian et al. 2001b, 2002). At this stage, 
and given the survey capability, it 
seems natural to search for TeV gamma ray emission not only 
from distant AGN, but to turn the instruments to our nearest
neighbor galaxy, M31. While -- as will be discussed below --
one would not expect conventional sources in M31, equivalent, e.g.,
to the Crab Nebula, to be visible over the distance, there is the
possibility of new classes of sources, which have not been discovered
in our own Galaxy because they are not particularly prominent in
other wavelength regimes, possibly since they
are heavily obscured by interstellar matter.
While it is hardly possible to survey with Cherenkov telescopes the
whole Galactic sky, it is relatively easy to cover 
all of M31 in searching for new TeV
sources.

In this work, we report on a TeV survey of M31 using the 
HEGRA IACT system. We will first briefly summarize the relevant
characteristics of M31 in the remainder of this section, and then
discuss the data analysis and the results.

Excellent summaries of the characteristics of M31 can be found
in (Hodge 1992) and in (Berkhuijsen et al., 2000). 
M31 is an early type spiral galaxy with a
prominent bulge; at a distance of about 770 kpc 
(Freedman \& Madore 1990, Stanek \& Garnavich 1998,
van den Bergh 2000) -- slightly increased compared to earlier
estimates (de Vaucouleurs 1972).
M31 is 30\% to 50\% larger than our own Galaxy
(Hodge 1992) and extends over
a few degrees on the sky. From rotation curves, the total mass
within a 24~kpc radius is estimated to $1.9 \cdot 10^{11}$~M$_\odot$
(Rubin \& Ford 1970, Evans \& Wilkinson 2000 and refs. given there), 
with a mass/luminosity ratio of $12 \pm 1$. The total mass of M31
could be as high as $1.2 \cdot 10^{12}$~M$_\odot$
(Evans \& Wilkinson 2000), indicative of a significant 
dark halo. Based on the study of the rotation curves within a few 
arc-seconds from the center, the presence of a central black
hole of $3 \cdot 10^7$~M$_\odot$ is inferred
(Kormendy \& Bender 1999 and refs. given there). 

A prominent feature of M31 is the concentration of the interstellar
medium in an annulus about 10~kpc from the center, with additional
spiral arm segments near 5~kpc and extending out to 20~kpc. The
ring structure is clearly evident in the optical, 
radio (e.g. Dame et al. 1993, Beck et al. 1998, Beck 2000) and
infrared (Xu \& Helou 1996) domains, and is also indicated in  the 
distribution of X-ray sources (Supper et al. 2001). 
Star formation in M31 is largely
concentrated in the annulus, where the cosmic ray rate is 
expected to be enhanced, also due to the exceptionally regular
toroidal magnet field associated with the structure, with a 
strength around 5~$\mu$G (Beck 1982, Berkhuijsen et al. 1993).

On the basis of CO (Dame et al. 1993, Berkhuijsen 1993) and 21 cm surveys
(Cram et al. 1980), the total mass of interstellar
gas is estimated to be similar for M31 and the Galaxy; however,
whereas in the Galaxy H2 and HI are found at roughly equal amounts,
the total H2 mass of M31 is estimated to about $3 \cdot 10^8$~M$_\odot$,
an order of magnitude below the mass in HI of about 
$4 \cdot 10^9$~M$_\odot$. The star formation rate in M31, in
the range of 0.2 to 0.5~M$_\odot$/y
(Xu \& Helou 1996, Walterbos 1988), is significantly smaller
than in the Milky Way (Mezger 1988). While optical searches for Supernova
remnants in M31 have resulted in a significant number of 
objects -- well over 200 
(Braun \& Walterbos 1993, Magnier et al. 1995) -- the Supernova rate is also
below that of our Galaxy. Braun \& Walterbos (1993) 
estimate a rate of 1.25/century for M31, and
Tamman et al. (1994) derive a rate of 1.2/century,
compared to about 2.5/century for our Galaxy (Dragicevich et al. 1999
and refs. given there); Berkhuijsen (1984) finds similar ratios.
However, the errors
associated with these numbers are large, and equal rates in both Galaxies
are not completely ruled out. The cosmic-ray electron flux in M31 has been
estimated from the intensity of the radio continuum using 
equipartition arguments (e.g.,
\"Ozel \& Berkhuijsen 1987), again resulting in a somewhat   
smaller density of relativistic electrons in M31.

M31 has been surveyed extensively in X-rays. First individual
sources were resolved in the Einstein survey
(van Speybroeck et al. 1979); including later
reanalysis of the data, 108 sources were found. Two ROSAT
surveys 
(Supper et al. 1997, 2001)
resulted in 560 sources, with X-ray luminosities
ranging from $4 \cdot 10^{35}$ to $4 \cdot 10^{38}$~erg/s. The
central region of M31 was more recently surveyed by XMM
(Shirey et al. 2001, Osborne et al. 2001)
and 116 sources were located in the inner 30'. Chandra
observation of the center region 
(Garcia et al. 2000) resolved the nuclear source
seen by Einstein and ROSAT in five discrete sources, one of 
which is most likely associated with the central black hole.

At gamma ray energies, an upper limit of $1.8 \cdot 10^{-8}$
photons/cm$^2$s above 100 MeV
was derived from EGRET data (Blom et al. 1999); 
earlier EGRET results were given by Sreekumar et al. (1994).
The limit implies
that the gamma ray flux of M31 is less than 1/2 of that of 
the Milky Way, which is not too surprising if cosmic ray 
production is indeed linked to Supernova activity. Earlier
upper limits on the TeV gamma-ray flux were given by
Cawley et al. (1985) at a flux of 
$2.2 \cdot 10^{-11}$ photons/cm$^2$ above 400 GeV.

On basis of these characteristics, we will address the
expectations concerning the visibility of TeV sources,
in relation to the sensitivity of the instruments, which have 
reached the range of $10^{-12}$/cm$^2$s for the flux above 1 TeV,
equivalent to a flux of about 5\% of the flux detected from the
Crab Nebula
\footnote{Here and in the following, we will use a flux
of $1.75 \cdot 10^{-11}$~cm$^{-2}$s$^{-1}$ above 1 TeV
for the Crab Nebula, taken from Aharonian et al. (2000).}.
This flux sensitivity implies that only sources in M31
with a TeV luminosity above a few $10^{38}$~erg/s have a chance
to be detected. The visibility of gamma-ray sources in M31
was so far mostly discussed in relation to the sub-GeV satellite
energy range, see e.g. 
\"Ozel \& Berkhuijsen (1986), \"Ozel \& Fichtel (1988), Pavlidou \& Fields (2001).

Known Galactic TeV point sources are have peak fluxes at most in the range
of the Crab flux. Scaling the flux
according to the distance, such sources in M31 generate a flux
around $10^{-6}$ to $10^{-5}$ Crab units, well below the detection
threshold of current instruments. 

Another potential source is the diffuse radiation generated in
interactions of cosmic rays with the interstellar medium. Optimistically
assuming the same emissivity as in our own Galaxy, the 
predicted TeV flux is about $10^{-12} M_6/d^2_{\mbox{\small kpc}}$ cm$^{-2}$s,
where $M_6$ is the target mass in units of $10^6$ solar masses, 
and $d_{\mbox{\small kpc}}$ the distance in kpc
(Aharonian, 2001a). Expressed in Crab units, the
expected flux is $6 \cdot 10^{-4}$, again below the sensitivity
threshold. Scaling with the lower SN rate in M31 reduced predictions
by another factor of a few (Pavlidou \& Fields 2001).

Like in AGN, the central black hole would be expected to generate
high-energy gamma rays when accreting ambient matter. The limiting
Eddington luminosity of the central black hole is $4 \cdot 10^{45}$~erg/s,
many orders of magnitude above the detection threshold. However,
while the X-ray luminosity of the center of M31 
-- about $10^{39}$ erg/s in the 1-10~keV range -- is more or less
in the range expected for normal galaxies, the radio emission is
unusually low, indicating only normal or even sub-normal
 activity in the core of M31 (Yi \& Boughn 1999).

A final speculative source might be the annihilation radiation
of neutralino dark matter accumulating near the center of M31
(see, e.g., Bergstr\"om et al., 1998, 1999 and 2001, Ellis et al, 2002).
In the framework of supersymmetric particles as the main
constituent of non-baryonic dark matter in the universe, both
monochromatic
gamma-ray line emission and continuum emission 
is expected from the self-annihilation of the 
lightest stable particle (LSP).
Predicted flux levels are quite uncertain, and depend on how
cuspy the distribution of dark matter near the center is --
an issue still under much debate. Fluxes currently estimated from
dark matter annihilation in our own Galactic center 
-- and uncertain by up to two orders of magnitude -- are at or
below the detection limits, and scale to fluxes below $10^{-4}$
Crab units from the center of M31. 
The flux due to neutralino annihilation from M31 in the CELESTE
energy range ($> 30$~GeV) was recently discussed by Falvard et al.
(2002), who argue in favor of a large dark matter component near
the center of M31, on the basis of rotation curves by Braun (1991).
Still, the signal is at best marginally detectable.
However, it has been speculated
(Gondolo \& Silk 1999) that under the influence of a central
black hole, as in our Galaxy or in M31, the dark matter 
distribution should develop a spike, enhancing the annihilation
rate by many orders of magnitude. On the other hand, it was 
argued by Merrit et al. (2002) that the spike is reduced due to
mergers of dark matter clusters, and by Ullio et al. (2001) that
a more detailed dynamical modeling shows only a much weakened
spike.

In summary, one must conclude that such ``conventional'' TeV
sources in M31 are most likely not detectable with current instruments.
On the other hand, as evidenced by the X-ray sources, there
are objects in M31 in the relevant $10^{38}$ erg/s luminosity
range; if the spectrum extends to highest energies, such sources
would be visible in a survey, providing enough motivation for
a measurement campaign with the HEGRA instrument.

\section{The HEGRA telescope system and the M31 data set}

The HEGRA system of five imaging atmospheric Cherenkov telescopes
is located on the Canarian Island of La Palma, at 2200 m asl.
Each of the telescopes has a 8.5~m$^2$ mirror and is equipped
with a 271-pixel photomultiplier camera with a $4.3^\circ$
field of view. Cherenkov images in at least two telescopes
are required to trigger the recording of event data;
the energy threshold of the instrument is about 500~GeV
for sources in the zenith. From the stereoscopic 
Cherenkov images, the direction of the primary gamma-ray is
reconstructed with an event-by-event precision of $0.1^\circ$.
Cosmic ray background showers are strongly suppressed on the
basis of image shapes. The effective detection area of the
instrument is determined by the extent of the Cherenkov light
pool and varies from $5 \cdot 10^3$~m$^2$ at the nominal threshold to
$10^5$~m$^2$ well above threshold. 
The detection rate for gamma rays is rather uniform
within the central $2^\circ$ of the field of view,
and drops to 63\% of its peak value at $1.8^\circ$ from the 
optical axis. Details about the telescope
system and the analysis techniques are given in 
(Aharonian et al. 1999, Konopelko et al. 1999)

M31 was observed in August, September and November 2001, using
all five telescopes. In order to cover all of M31,
observation time was distributed between
three tracking positions, one centered on the core of
M31 at RA 0h 42' 44" DEC $41^\circ$~16'~9.12", one 
displaced by $0.56^\circ$ to the SW at
RA 0h 40' 30" DEC $40^\circ$~39'~0" and one 
displaced to the NE at RA 0h 44' 43" DEC
$41^\circ$~53'~0". After data cleaning, 20.1~h of
good observation time were selected, most of which were
taken at Zenith angles below $25^\circ$. For calibration and
reference, 9.7~h of Crab Nebula data taken in October and
November were selected.

\section{Data analysis and search for TeV point sources}

\begin{figure}
\begin{center}
\mbox{
\epsfxsize7.15cm
\epsffile{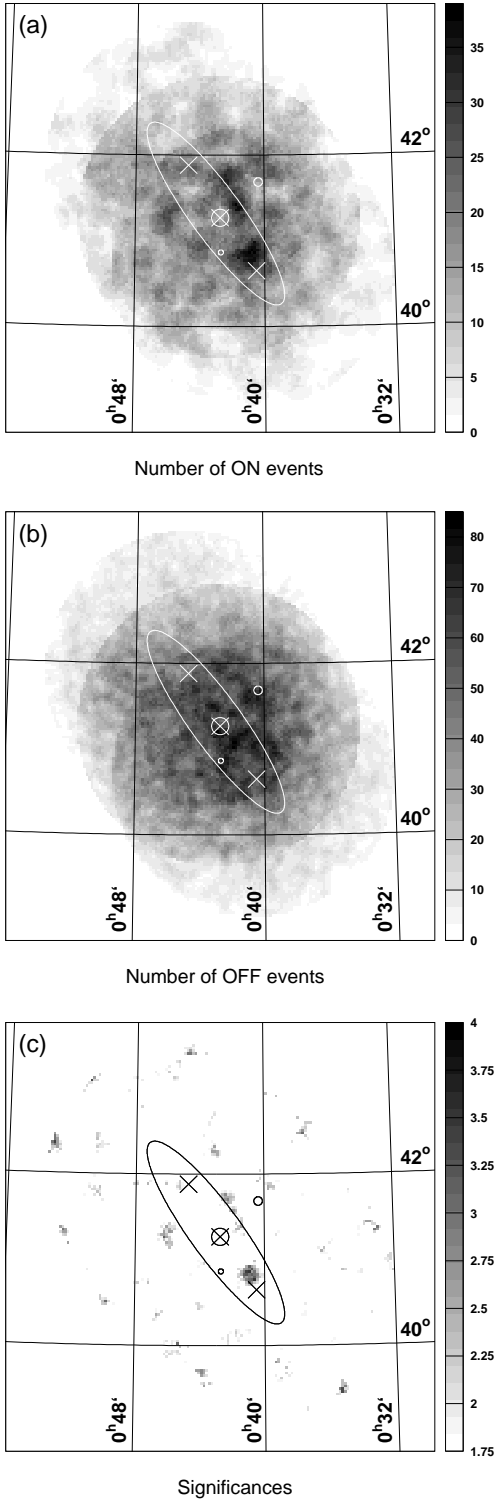}}
\end{center}
\caption{(a) Number of gamma ray candidates for each
point of the search grid, (b) estimated number of
background events, (c) significance for an excess of 
signal events.
For reference, the three tracking locations are indicated by crosses,
the locations of the center of M31 and
of the 10~kpc dust ring are indicated. The small circles indicate the
positions of M32 and NGC205.}
\label{fig_events}
\end{figure}

\begin{figure}
\begin{center}
\mbox{
\epsfxsize8.0cm
\epsffile{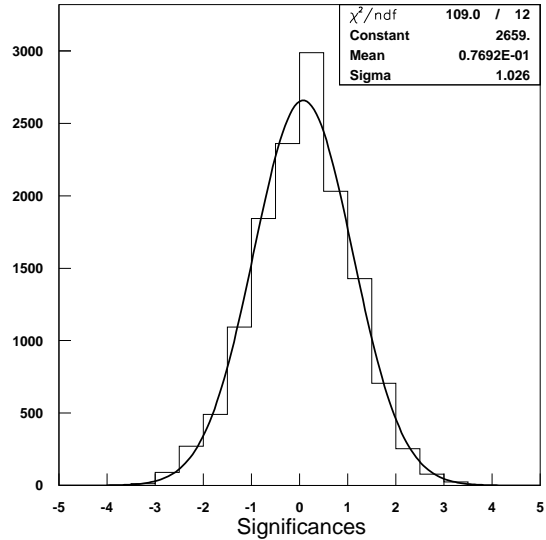}}
\end{center}
\caption{Distribution of the significance for an excess
at the points of the search grid. We note that the significances
derived for adjacent points are correlated, since the spacing
of points is of the same scale as the experimental resolution.
There is no point above 3.5$\sigma$ and
the distribution is well represented by a Gaussian of mean 0.08 and
width 1.026.}
\label{fig_sigdis}
\end{figure}

\begin{figure}
\begin{center}
\mbox{
\epsfxsize8.0cm
\epsffile{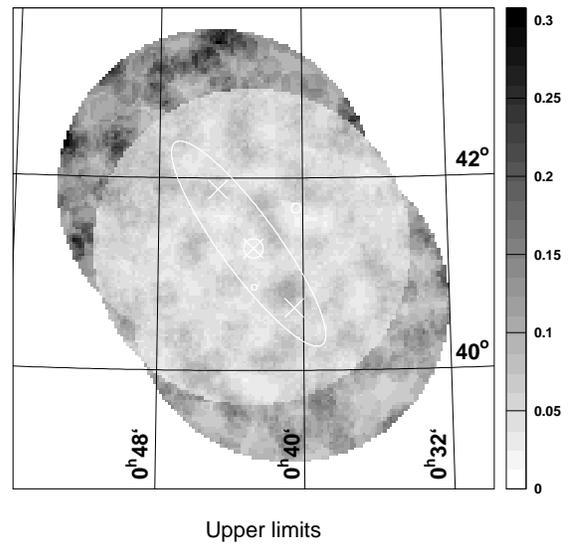}}
\end{center}
\caption{Upper limits on the flux from TeV point 
sources in M31, expressed in units of the Crab flux.
For reference, the locations of the center of M31,
the 10~kpc dust ring and the two companion galaxies M32 and NGC205 
are indicated. Tracking positions are indicated by crosses}
\label{fig_limits}
\end{figure}

The data analysis follows to a large extent the procedures 
developed for earlier surveys (Aharonian et al. 2001b, 2002), and concentrates
on the search for sources which appear pointlike on the
$0.1^\circ$ scale of the angular resolution of the instrument.
The search for sources is carried out on a $6^\circ$ by $6^\circ$ grid
with a spacing $0.031^\circ$ well below the angular resolution.
For optimal rejection of cosmic-ray backgrounds, only
events with at least four Cherenkov images (out of a maximum of five)
were used, and gamma-ray candidates were selected by requiring
on the mean scaled with $\bar{w}$ of images below 1.1; gamma rays generate
a peak in $\bar{w}$ at 1.0, with width 0.1, whereas the distribution
for the more diffuse cosmic ray images peaks at $\approx 1.6$
(see Fig. 1 of Aharonian et al., 2001b).
For each grid point, gamma ray candidates with reconstructed directions
within $0.143^\circ$, consistent within the angular resolution, 
were counted. Here, shower directions up to $1.8^\circ$ from
the optical axis of the telescope system were accepted.
Since no specific off-source data were taken,
the expected background for each grid point is determined from
the data itself, making use of the fact that the detection rate
of the telescope system is (a) virtually flat near the center
of the field of view, and (b) has an azimuthal symmetry with 
respect to the optical axis. Two methods for background estimation
were applied: For search points more than 
$0.143^\circ$ from the optical axis, three regions the same size
as the signal region, but rotated by $90^\circ$, $180^\circ$
and $270^\circ$ in azimuth with respect to the optical axis
were used to estimate the background. This procedure cannot be
used for search regions near the optical axis, since signal
region and background regions overlap. Instead, a circular
background region was chosen, centered on the signal region,
with an inner radius of twice the radius of the signal region,
and an outer radius adjusted such that the area of the background
region is three times the area of the signal region. A cross-check
of both methods in the regions of overlap indicates only negligible systematic
differences between both methods. In the following we therefore give
results for the ring model.
Fig.~\ref{fig_events}(a) shows the distribution of detected gamma-ray
candidates from M31, selected by a $\bar{w} < 1.1$ cut. 
Fig.~\ref{fig_events}(b) gives the predicted number of background
events. No obvious excess of signal events is visible at
any grid point. The significance for each grid point is then
calculated based on Li \& Ma (1983), and is shown in Fig.~\ref{fig_events}(c).
The distribution of significances for all grid point 
approximates a Gaussian distribution with unit width, as expected
in the absence of a genuine signal (Fig.~\ref{fig_sigdis});
the point with the highest significance of 3.5 $\sigma$
is well compatible with the expected tail of the Gaussian
distribution, and should be interpreted as an upward
fluctuation.

We conclude that there is no statistically significant 
indication of a TeV gamma ray point source in M31, and
proceed to extract upper limits for the source flux.
To derive the limits, first the 99\% confidence level
on the number of excess events is calculated according 
to Helene (1983).
 The number is then corrected for the 28\% acceptance loss of the
angular cut of $0.143^\circ$ relative to the direction of the source.
The number of upper limit counts is normalized to the expected
rate detected from the
Crab nebula at the relevant zenith angle; since the
Crab observations were carried out with the source positioned
at a fixed angle of $0.5^\circ$ from the optical axis, 
the dependence of detection rate on the position within
the field of view was corrected based on Monte Carlo
simulations (see Fig. 3 of Aharonian et al., 2001b).
The limits are given in units of the Crab flux and 
refer to a characteristic energy of 1~TeV. The limits
are calculated assuming the Crab energy spectrum, but
are relatively insensitive to this assumption; a 
variation in spectral index of $\pm 0.5$ changes the
flux limits by less than 10\%.

The resulting flux limits are shown in 
Fig.~\ref{fig_limits}, and range from 3.3\% of the
Crab flux for the center of M31 to about 30\% at the periphery.
For two other objects in the survey range, M32 and NGC205,
limits of 4.4\% and 2.8\% of the Crab flux are derived,
respectively.

\section{Search for supersymmetric dark matter in M31}

Beyond the general survey,
the good energy resolution of the HEGRA telescope system 
allows a dedicated search for supersymmetric dark matter in M31, 
looking for the line emission from the self-annihilation of the 
lightest stable particle (LSP). We have searched for line emission in the energy spectrum 
from the central part of M31 ($r<1.4$~kpc), and compare the results
with predictions of  the 
flux of gamma-rays from the annihilation of
neutralinos ($\chi_0$) in the framework of the minimal supersymmetric
standard model (MSSM) using 
a spatial distribution of the radial density of neutralinos in M31
in concordance with the measured rotation curve 
(Braun  1991, Falvard et al. 2002).

   The event selection and background estimate for the spectral analysis deviates
from the approach taken for the survey for point-sources in M31. Events with
three or more Cherenkov images were used. In order to reduce the
background from isotropic cosmic ray events, a tight cut on the direction has been chosen 
for  the central part of M31. Only events from within a cone of radius $0.105^\circ$
are accepted. At a distance of 770~kpc this
corresponds to the inner $1.4$~kpc of M31. 
The cut on $\bar{w}$  remains the same ($\bar{w}<1.1$).  
The sensitivity for line emission depends on the energy resolution of the detector. 
For this analysis, an improved energy reconstruction algorithm as described in Hofmann et al. 
(2000)  has been applied. The relative energy resolution ($\Delta E/E$) reaches 
10\,\% for a wide energy band from the threshold of 500~GeV up to 10~TeV. 
The search bin in energy is $12~\%$ wide,
concentrating on the central section of a monoenergetic line in order to achieve best
signal to noise ratios. We have
not attempted to optimize the search bins individually for the different energy regions;
the optimal choice would have
to take the diminishing event statistics at higher energies into account.   
The energy bins are equally spaced on a logarithmic scale such that the bin centers
are separated by 0.025 on a decadic logarithmic scale making the bins correlated. 

\begin{figure}
\begin{center}
\mbox{
\epsfxsize9.0cm
\epsffile{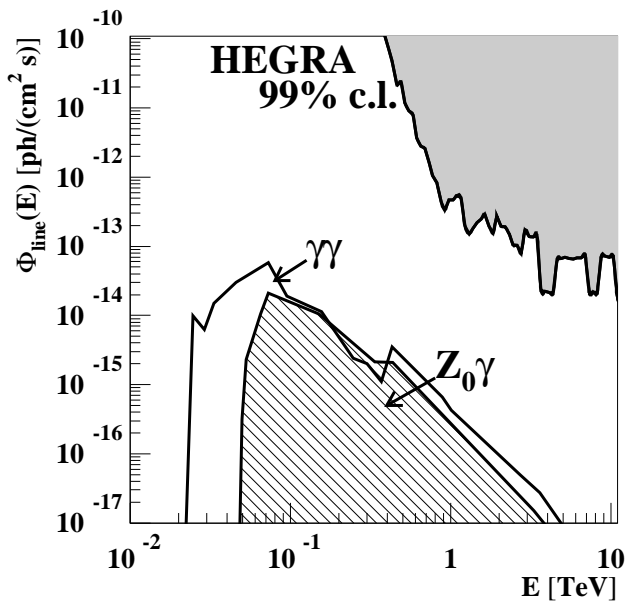}}
\end{center}
\caption{ Exclusion region (grey shaded region) 
for the the gamma-ray line
flux from the center of M31. The two sets
of curves labeled $\gamma\gamma$ and $Z^0\gamma$ indicate the
upper range of model predictions for the minimal supersymmetric
models calculated in Bergstr\"om, Ullio, and Buckley (1998). The
dark matter halo model for M31 has been taken from Falvard et al. (2002), 
Table 1, using a mass-to-light ratio of 3.5 for the bulge and 
2.5 for the disk.}
\label{fig_dm}
\end{figure}

The background expectation of the measurement 
is determined using seven independent OFF-regions with similar acceptance to the
ON-region. For the calculation of upper limits on the number of excess events $N_{\gamma}^{99\,\%}$,
the 99\,\% c.l. upper limits were calucated 
according to Helene (1983), and an upper limit on the
rate from a $\gamma$-ray line was derived.
The corresponding flux limits 
were calculated using collection areas $A_{eff}(E,\theta)$ 
derived from Monte Carlo simulations (Konopelko et al. 1999) applying the same
reconstruction methods and event selection as for the data analysis. The resulting 
exclusion region is indicated in Fig.~\ref{fig_dm} as the grey shaded region.

A prediction for the line flux emission was derived using scans of 
the 7-parameter space of MSSM performed by Bergstr\"om et al. (1998). 
The expected flux 
is given by Eq. (6) of Falvard et al. (2002) or the equivalent 
Eq. (13) of Bergstr\"om et al. (1998) (scaled to the distance to M31):
\begin{eqnarray*}
I_\gamma = & (3.18 \cdot 10^{-13} \mbox{ photons cm$^{-2}$ s$^{-1}$}) \\
&
\times~\left( {\langle \sigma v \rangle N_\gamma \over 10^{-25} 
\mbox{ cm$^3$ s$^{-1}$}} \right)
\left( {500 \mbox{ GeV} \over m_\chi} \right)^2 \Sigma_{19}
\end{eqnarray*}
Apart from trivial factors, it
depends on the velocity-weighted annihilation
cross section $\langle \sigma v\rangle$ and on the line-of-sight
and solid-angle integrated squared LSP mass density $\rho$
$$
\Sigma = \int \int \rho^2~\mbox{d}s~\mbox{d}\Omega  
$$
written as $\Sigma_{19}$ in units of
$10^{19}$~GeV$^2$~cm$^{-5}$.
The
envelope of the allowed $\langle \sigma v\rangle$ vs. $M_{\chi}$ 
space was used as given in
Figs. 1 and 2 of Bergstr\"om et al. (1998). 
The unitless parameter $\Sigma_{19} \approx 1.7$
corresponds to $\Sigma_{19}=3$ of  Falvard et al. (2002),
scaled to the inner 1.4 kpc on the basis of the $\rho(r)$
distributions of Fig. 3(b) of Falvard et al. (2002)
and using a distance of 770 kpc.
This value represents the upper range of the different models
for the dark matter halo of M31 discussed in this reference.
The predicted line flux is 
given in Fig.~\ref{fig_dm}. 
We have indicated the two  individual final states with line emission ($\chi^0\chi^0\rightarrow
\gamma\gamma$ and $\chi^0\chi^0\rightarrow Z^0\gamma$) in the figure. The model prediction given here 
is based upon smoothly 
distributed dark matter. Within this model, no signal from M31 with the current sensitivity of 
Cherenkov detectors is to be expected. Very favorable conditions as for example
clumpiness of dark matter would lead to a considerable increase of the annihilation rate
(Bergstr\"om et al. 1999,Taylor \& Silk 2002). It is conceivable 
that the flux level of some models would be become detectable by tuning the distribution of the
dark matter halo and by invoking mechanisms to allow for larger cross sections 
e.g. quintessence (Salati 2002) or
anomaly-mediated supersymmetry breaking scenarios (Ullio 2001). However, 
in the absence of a signal, no exotic speculations are required.

\section{Concluding remarks}

For the first time, the Andromeda galaxy was surveyed
for point-like sources of TeV gamma rays. No sources 
were detected at levels exceeding a few percent of the Crab
flux, equivalent to a source strength of about 
$3 \cdot 10^{38}$ erg/s above 1 TeV. The fact that limits
within of a few percent of the Crab flux were achieved with
only 20~h of observation time distributed over three 
locations indicates the enormous progress of the 
detection technique during the last years.
While TeV sources of a detectable strength are well
within the range of source energetics,
the lack of such sources does not come as a big surprise,
given the discussion presented in the introduction and the
comparison with Galactic TeV sources.
The observations rule out
a new and hitherto undiscovered population of very
strong VHE sources in normal galaxies.

 We have specifically searched for the first time for a gamma-ray line from M31 with
an energy resolution of $\Delta E/E=10\,\%$. We have not found any indication for a signal, setting
upper limits at the level of $10^{-13}$cm$^{-2}$s$^{-1}$ above 1~TeV. The 
expectation for the $\gamma$-ray line flux from M31 is by 2 orders of magnitude lower than the sensitivity reached
with the current instruments. However, with a lower energy threshold (100 GeV) and better sensitivity 
combined with
a similar energy resolution as achieved by next-generation experiments
like H.E.S.S. and VERITAS it will be possible to probe the MSSM parameter space for the Galactic center and 
nearby galaxies.

\section*{Acknowledgements}

The support of the HEGRA experiment by the German Ministry for Education
and Research BMBF and by the Spanish Research Council
CICYT is acknowledged. We are grateful to the Instituto
de Astrof\'\i sica de Canarias for the use of the site and
for providing excellent working conditions. We gratefully
acknowledge the technical support staff of Heidelberg,
Kiel, and Munich.

\end{document}